\documentclass[11pt]{article}

\usepackage[utf8]{inputenc}
\usepackage[T1]{fontenc}
\usepackage{lmodern}
\usepackage{microtype}
\usepackage{graphicx}
\usepackage{booktabs}
\usepackage{multirow}
\usepackage{amsmath, amssymb}
\usepackage{url}
\usepackage{enumitem}
\usepackage{xcolor}

\usepackage{float}
\usepackage{placeins}

\usepackage{array}
\usepackage{tabularx}
\usepackage{ragged2e}
\usepackage{adjustbox}
\usepackage{tabularray}
\usepackage{booktabs}
\usepackage{makecell}
\UseTblrLibrary{booktabs}
\newcolumntype{L}[1]{>{\RaggedRight\arraybackslash\hspace{0pt}}p{#1}}
\newcolumntype{C}[1]{>{\Centering\arraybackslash\hspace{0pt}}p{#1}}
\newcolumntype{Y}{>{\RaggedRight\arraybackslash\hspace{0pt}}X}

\usepackage[nottoc]{tocbibind}

\usepackage{hyperref}
\hypersetup{
  colorlinks=true,
  linkcolor=blue,
  citecolor=blue,
  urlcolor=blue,
  hypertexnames=false
}

\title{Generative AI for Quantum Circuits and Quantum Code:\\ A Technical Review and Taxonomy}
\author{
  Juhani Merilehto \\
  University of Vaasa \& University of Turku \\
  \texttt{merilehto@pm.me}
}
\date{}

\begin{document}
\maketitle

\begin{abstract}
We review thirteen generative systems and five supporting datasets for quantum circuit and quantum code generation, identified through a structured scoping review of Hugging Face, arXiv, and provenance tracing (January--February 2026). We organize the field along two axes---artifact type (Qiskit code, OpenQASM programs, circuit graphs) crossed with training regime (supervised fine-tuning, verifier-in-the-loop RL, diffusion/graph generation, agentic optimization)---and systematically apply a three-layer evaluation framework covering syntactic validity, semantic correctness, and hardware executability. The central finding is that while all reviewed systems address syntax and most address semantics to some degree, \emph{none} reports end-to-end evaluation on quantum hardware (Layer~3b), leaving a significant gap between generated circuits and practical deployment. \emph{Scope note}: ``quantum code'' refers throughout to quantum \emph{program} artifacts (QASM, Qiskit); we do not cover generation of quantum error-correcting codes (QEC).
\end{abstract}

\section{Introduction}\label{sec:intro}

Generative AI for quantum software has diversified from quantum-aware code assistants into multiple technical families that synthesize quantum artifacts at different abstraction levels. The important axis of differentiation across these systems is not ``LLM vs.\ non-LLM,'' but \emph{how semantic correctness is defined and enforced}: unit tests, fidelity proxies, objective-function scores, or entanglement proxies. This review imposes structure on this fragmented landscape.

\paragraph{Scope.}
We focus on \emph{generative} systems that output quantum artifacts intended to be executed or compiled: (i)~quantum circuits as gate sequences or graphs; (ii)~OpenQASM (2.0 and 3.0) programs; and (iii)~Qiskit (Python) code that constructs circuits. We exclude systems where quantum circuits are internal components but outputs are non-circuit data (``quantum-enhanced'' generative modelling). We use the following terminology throughout:
\begin{itemize}[leftmargin=*,nosep]
  \item \textbf{Syntactic validity}: the output parses/compiles under the target grammar/toolchain.
  \item \textbf{Semantic correctness}: the generated artifact implements the intended unitary, algorithm, or task objective.
  \item \textbf{Hardware executability}: the artifact transpiles and runs under realistic device constraints (connectivity, gate set, noise) with acceptable resource usage.
\end{itemize}

\paragraph{OpenQASM 2.0 versus 3.0.}
Several reviewed systems target OpenQASM 2.0 \cite{openqasm2} while others target OpenQASM 3.0 \cite{openqasm3}. OpenQASM 2.0 is a straight-line gate-sequence language; 3.0 introduces classical control flow (\texttt{for}, \texttt{while}, \texttt{if-else}), typed variables, subroutine definitions, and timing instructions. For generative models, this distinction matters in three ways: (1)~the grammar space is considerably larger, increasing the probability of syntactically invalid output; (2)~semantic correctness becomes harder to verify because classical control flow creates path-dependent behaviour; and (3)~generated circuits may exploit features (e.g., mid-circuit measurement and feed-forward) that current simulators and hardware support unevenly. Table~\ref{tab:models} identifies which QASM version each system targets; systems operating on 2.0 and 3.0 are \emph{not} directly comparable in generation difficulty or evaluation complexity.

\paragraph{Positioning against classical code generation.}
Classical code LLMs such as Codex \cite{codex}, AlphaCode \cite{alphacode}, and CodeBERT \cite{codebert} generate programs evaluated primarily via unit tests and execution-based feedback. Unit-test evaluation transfers directly to Qiskit code generation, as demonstrated by QiskitHumanEval \cite{qiskit_humaneval}. However, quantum semantic equivalence checking---verifying that two circuits implement the same unitary---is fundamentally more expensive: statevector simulation requires $O(2^n)$ memory for $n$ qubits, and full unitary comparison costs $O(4^n)$. No general polynomial-time equivalence checker is known for arbitrary unitaries at scale. Hardware executability as a first-class constraint---connectivity maps, native gate sets, and coherence-time budgets---has no classical analogue. These differences motivate the three-layer framework developed in \S\ref{sec:eval_stack}.\looseness=-1

\section{Review Methodology}\label{sec:methods}

\subsection{Search and Screening}
We conducted a scoping review following a structured search protocol modelled on PRISMA-ScR reporting guidelines. All screening and inclusion decisions were performed by a single reviewer (the author); no independent second screening was conducted. This is acknowledged as a limitation in \S\ref{sec:threats}. Sources were assembled between January~1 and February~15, 2026 via three channels:
\begin{enumerate}[leftmargin=*]
  \item \textbf{Model-hub search} on Hugging Face: four keyword queries against model-card content---\texttt{"QASM"} (11~hits), \texttt{"quantum circuit"} (40~hits), \texttt{"OpenQASM"} (3~hits), \texttt{"Qiskit generator"} (0~hits)---yielding 35 unique base model cards after deduplication and collapsing quantized redistributions.
  \item \textbf{Paper search} on arXiv (categories \texttt{cs.AI}, \texttt{cs.LG}, \texttt{quant-ph}; submissions $\leq$ 2026-02-15): five keyword queries yielding 193 unique papers after cross-query deduplication. The largest result set (185 papers for RL~$+$~quantum circuit) is dominated by quantum-enhanced RL works that do not \emph{generate} circuits; these were excluded during screening.
  \item \textbf{Provenance follow-up} via GitHub repositories, Hugging Face organization pages, and backward/forward citation tracing. This channel recovered three systems (Granite-3.2-8b-Qiskit, Qwen2.5-14B-Qiskit, KetGPT) whose model cards did not match keyword queries.
\end{enumerate}

\paragraph{Screening flow.}
The combined pool of 228 unique candidates (35 HF model cards $+$ 193 arXiv papers) was screened in two stages: (i)~title/abstract screening removed 190 candidates (172 quantum-enhanced RL/VQC papers $+$ 18 non-generative HF model cards); (ii)~full-text screening of the remaining 38 candidates removed 16 (11 for insufficient technical disclosure, 5 for producing outputs outside scope). After deduplication across channels, 13 generative systems and 5 datasets were retained.

\paragraph{Inclusion criteria.}
A system was included if public artifacts (paper, model card, or repository) jointly disclosed at least two of: (i)~model architecture or parameter count, (ii)~training data source and approximate scale, (iii)~at least one quantitative evaluation metric. Systems with partial disclosure were included and annotated with ``Unspecified in source'' for missing fields. Fully closed systems with no public disclosure were excluded.

\paragraph{Treatment of partial disclosure.}
``Insufficient technical disclosure'' was applied when a system's public artifacts did not meet the two-of-three criterion above and the missing information could not be inferred from the repository. Systems excluded for this reason are noted in footnotes rather than listed individually, as their omission reflects disclosure limitations rather than technical inadequacy.

\section{Background and Timeline}\label{sec:timeline}

The systems reviewed here belong to a broader continuum of automated quantum circuit construction. Before the current wave of generative-model-based approaches, the field developed substantial foundations in evolutionary and reinforcement-learning-based circuit synthesis. Genetic algorithms have been applied to quantum circuit compilation since at least 2019 \cite{rasconi2019genetic}, and deep RL was demonstrated for quantum compiling by Moro et al.\ \cite{moro2021rl_compiling}. Multi-objective evolutionary architecture search \cite{lu2023qasbench} further matured the space. These pre-LLM methods typically operate by sequential gate placement guided by heuristic or learned value functions, and remain competitive for structured synthesis tasks. The generative-model wave reviewed here (2024--2026) differs primarily in its use of large pre-trained language or diffusion models and in its ambition to generalize across task families rather than optimize for a single target unitary.

Table~\ref{tab:timeline} provides a chronological overview. The field has progressed from benchmark and dataset construction (2020--2024) through supervised generation models (2024--2026), with verifier-in-the-loop and agentic systems emerging prominently in 2025.

\begin{table}[!htbp]
\centering
\scriptsize
\setlength{\tabcolsep}{2.5pt}
\renewcommand{\arraystretch}{1.05}
\begin{tabularx}{\textwidth}{%
  @{}%
  >{\RaggedRight\arraybackslash}p{0.9cm}%
  >{\RaggedRight\arraybackslash}p{2.8cm}%
  >{\RaggedRight\arraybackslash}p{1.4cm}%
  >{\RaggedRight\arraybackslash}X%
  @{}}
\toprule
\textbf{Year} & \textbf{System / Dataset} & \textbf{Type} & \textbf{Key Innovation} \\
\midrule
2020 & QASMBench \cite{qasmbench_pnnl,qasmbench_repo} & Benchmark & Curated low-level OpenQASM 2.0 benchmark suite for NISQ evaluation \\[1pt]
2024 & genQC \cite{genqc_paper,gh_genqc} & Model & First diffusion model for quantum circuit synthesis; text-conditioned denoising over discrete gate tokens \\[1pt]
2024 & KetGPT \cite{ketgpt_paper,gh_ketgpt} & Model & GPT-based transformer for generating realistic OpenQASM 2.0 circuits (dataset augmentation) \\[1pt]
2024 & AltGraph \cite{altgraph_paper} & Model & Generative graph models (D-VAE, DeepGMG) for circuit DAG rewriting and optimization \\[1pt]
2024 & QiskitHumanEval \cite{qiskit_humaneval} & Benchmark & Unit-test benchmark (101 tasks) for Qiskit code generation \\[1pt]
2024 & quantum-circuits-8k \cite{hf_quantum_circuits_8k} & Dataset & Synthetic text$\rightarrow$QASM 2.0 pairs with paraphrase augmentation \\[1pt]
2024 & QuantumLLMInstruct \cite{quantumllminstruct_paper,hf_quantumllminstruct} & Dataset & 500k+ claimed instruction-tuning pairs across 90+ quantum domains \\[1pt]
2024 & QCircuitBench \cite{qcircuitbench_paper} & Benchmark & 120k+ algorithm-design instances with verification oracles spanning 25 algorithms \\[1pt]
2025 & Granite-3.2-8b / Qwen2.5-14B Qiskit \cite{hf_granite_qiskit,hf_qwen_qiskit,dupuis2025} & Model & Industrial Qiskit code LLMs with GRPO post-training using quantum verifiable rewards \\[1pt]
2025 & UDiTQC \cite{uditqc_paper} & Model & U-Net-style diffusion transformer; outperforms genQC on entanglement and compilation \\[1pt]
2025 & Agent-Q (SFT) \cite{agentq_paper} & Model & SFT on 14k optimization circuits in OpenQASM 3.0 \\[1pt]
2025 & Barta et al.\ \cite{barta2025diffusion} & Model & Diffusion for parameterized quantum circuits; extends to continuous gate parameters \\[1pt]
2025 & QUASAR (SFT+RL) \cite{quasar_paper} & Model & Agentic RL with hierarchical 4-level reward; tool-augmented LLM \\[1pt]
2025 & Q-Fusion \cite{qfusion_paper} & Model & LayerDAG-based diffusion over circuit DAGs; 100\% syntactic validity in tested regimes \\[1pt]
2025 & genQC v2 \cite{genqc_v2} & Model & Multimodal diffusion generating discrete structure and continuous parameters simultaneously \\[1pt]
2025 & QAgent \cite{qagent_paper} & Model & Multi-agent LLM for autonomous OpenQASM programming; RAG + CoT + tool augmentation \\[1pt]
2025 & graph-data-quantum-rl \cite{hf_graph_data_quantum_rl} & Dataset & 14.5k rows with prompts, graphs, Hamiltonians, OpenQASM 3.0 circuits \\[1pt]
2026 & QuantumGPT-124M \cite{hf_quantumgpt_124m} & Model & Small specialist GPT-2 for OpenQASM 2.0; task-specific tiny LM feasibility \\
\bottomrule
\end{tabularx}
\caption{Chronological milestones in generative AI for quantum circuits and code. Years reflect first public appearance (preprint, model card, or repository).}
\label{tab:timeline}
\end{table}

\FloatBarrier

\section{Taxonomy of Generative Systems}\label{sec:taxonomy}

We organize reviewed systems along two axes: \textbf{artifact type} (Qiskit code vs.\ QASM vs.\ circuit graph) crossed with \textbf{training regime} (static SFT, verifier-in-the-loop RL, diffusion/graph generation, agentic optimization). This pair maximises separation among reviewed systems and aligns with the two practical questions a practitioner faces: ``What do I want the system to output?'' and ``How is correctness enforced during training?'' Alternative axis choices---qubit regime or verification cost---were considered but rejected as either degenerate (most systems operate in the small-qubit regime) or conflating distinct model designs that share a cost profile.

The six families are:
\begin{itemize}[leftmargin=*,nosep]
  \item \textbf{Qiskit code assistants}: general code LLMs adapted to Qiskit APIs, evaluated by executable unit tests.
  \item \textbf{OpenQASM generators (static SFT)}: supervised fine-tuned LMs producing OpenQASM for specific domains.
  \item \textbf{Specialist small LMs}: small models ($\sim$100M parameters) trained on text$\rightarrow$QASM instruction pairs.
  \item \textbf{Verifier-in-the-loop alignment}: RL/preference optimization with simulator-based rewards.
  \item \textbf{Graph and diffusion generators}: models operating on circuit DAGs or discrete tokenizations of gates and parameters.
  \item \textbf{Agentic systems}: multi-step generation with external tools (simulators, compilers) used for scoring and iterative improvement.
\end{itemize}

\subsection{Qiskit Code Assistants}
Granite-3.2-8b-Qiskit \cite{hf_granite_qiskit} and Qwen2.5-Coder-14B-Qiskit \cite{hf_qwen_qiskit} are general-purpose code LLMs extended pre-trained on a curated Qiskit corpus (approximately 50M tokens of Qiskit v2.0 API code) and fine-tuned with supervised instruction tuning. Evaluation uses QiskitHumanEval \cite{qiskit_humaneval}, a benchmark of 101 tasks where the metric is \textbf{pass@$k$}: the probability that at least one of $k$ generated completions passes all unit tests. More recent work \cite{dupuis2025} adds \textbf{GRPO} (Group Relative Policy Optimization) \cite{grpo_deepseekmath} post-training with quantum verifiable rewards. GRPO eliminates the critic network by estimating advantages relative to the group mean of sampled completions, reducing memory overhead. In the quantum setting, the reward function checks both syntactic correctness and functional equivalence via Qiskit Aer simulation.

\subsection{OpenQASM Generators and Specialist Small LMs}
Agent-Q \cite{agentq_paper,hf_agentq_sft} is a Qwen-based model fine-tuned on approximately 14,000 parameterized optimization circuits (QAOA, VQE, adaptive VQE) in OpenQASM 3.0. The released Hugging Face checkpoint is 4B parameters, though the paper does not clearly specify the base-model size. Evaluation measures \textbf{objective alignment}: Jensen--Shannon divergence between the output distribution of the generated circuit and the ground-truth distribution, as well as expectation-value discrepancy under problem-specific cost Hamiltonians.

QuantumGPT-124M \cite{hf_quantumgpt_124m,hf_quantum_circuits_8k} is a GPT-2-scale (124M-parameter) model trained on approximately 8,000 synthetic text$\rightarrow$OpenQASM 2.0 pairs with paraphrase augmentation. It targets small circuits ($\leq$5 qubits) and evaluates syntactic validity via parser checks and approximate task-type success via manual inspection.

\subsection{Verifier-in-the-Loop Alignment}
QUASAR \cite{quasar_paper,hf_quasar_rl} extends Agent-Q's SFT foundation with agentic reinforcement learning using GRPO. The key innovation is a hierarchical four-level reward computed by an external quantum simulation tool: (1)~a \textbf{syntax reward} for successful OpenQASM 3.0 parsing; (2)~a \textbf{distributional alignment} term (Jensen--Shannon divergence); (3)~an \textbf{expectation-value alignment} term comparing cost-Hamiltonian expectation values; and (4)~an \textbf{optimization usability} term assessing whether the generated circuit converges efficiently under further classical parameter optimization. The model interacts with a quantum tool server via HTTP, receiving structured feedback at each RL step.

\subsection{Graph and Diffusion Generators}
genQC \cite{genqc_paper,gh_genqc,hf_genqc_unitary} employs a \textbf{denoising diffusion model} on discrete circuit tokens. Circuits are represented as 2D tensors (rows = qubits, columns = time steps, cells = gate identities). The reverse process uses a conditional U-Net with text conditioning via frozen OpenCLIP embeddings. Evaluation uses \textbf{process fidelity} ($F = |\text{Tr}(U_{\text{gen}}^\dagger U_{\text{target}})|^2 / d^2$) and compilation success rate (typically 3--5 qubits). Model size is not reported as a single count due to the U-Net $+$ frozen CLIP architecture.

AltGraph \cite{altgraph_paper} uses three generative graph models---D-VAE (GRU and GCN variants) and DeepGMG---to transform quantum circuit DAGs. The models learn a latent space from which perturbations produce functionally equivalent circuits with reduced depth and gate count. Evaluation measures density-matrix MSE (0.0074 average) and post-transpilation gate count and depth reduction (37.55\% and 37.75\%). Model sizes are \textbf{unspecified in source}.

Q-Fusion \cite{qfusion_paper} adapts the LayerDAG diffusion framework to quantum circuit DAGs. It reports 100\% syntactic validity in tested regimes (small random circuits), though semantic evaluation beyond validity is limited. Model size is \textbf{unspecified in source}.

UDiTQC \cite{uditqc_paper} replaces genQC's U-Net backbone with a U-Net-style Diffusion Transformer (UDiT) combining multi-scale feature extraction with global self-attention. Evaluated on entanglement generation and unitary compilation (up to 8 qubits), it reports higher accuracy than genQC. The framework supports masked circuit editing and constrained generation. Model size is \textbf{unspecified in source}.

Barta et al.\ \cite{barta2025diffusion} extend diffusion to \emph{parameterized} quantum circuits, generating both discrete gate structure and continuous rotation angles---addressing a limitation of earlier discrete-token diffusion models. Accepted at QCE~2025. Model size is \textbf{unspecified in source}.

genQC~v2 \cite{genqc_v2} introduces a \emph{multimodal} denoising diffusion model that simultaneously generates circuit structure and continuous parameters using two independent noise processes with a shared conditioning mechanism. Model size is \textbf{unspecified in source}; evaluation disclosure in the public preprint is limited.

\subsection{Agentic Systems}
QAgent \cite{qagent_paper} is a multi-agent LLM system for autonomous OpenQASM programming. Given a natural language task description, it decomposes into sub-tasks dispatched to a \emph{Dynamic-few-shot Coder} (in-context learning for regular circuits) and a \emph{Tools-augmented Coder} (simulation tools for complex parameterized tasks). Both incorporate multi-round self-reflection with chain-of-thought reasoning and RAG. The system reports 71.6\% improvement over baseline LLMs on OpenQASM generation. Unlike QUASAR, QAgent uses prompt engineering and tool augmentation over a frozen base LLM rather than fine-tuning or RL. Model size depends on the pluggable base LLM.

\subsection{Dataset Augmentation Models}
KetGPT \cite{ketgpt_paper,gh_ketgpt} uses a GPT-based transformer to generate synthetic OpenQASM 2.0 circuits trained on algorithm-derived circuits from MQTBench. Its purpose is \emph{dataset augmentation} rather than task-directed generation: a three-fold verification process (manual inspection, transformer-based real-vs-random classification, and structural analysis) validates that generated circuits resemble real algorithm-based circuits. Model size is \textbf{unspecified in source}.

\subsection{Model Comparison}\label{sec:model_table}
Table~\ref{tab:models} summarizes the reviewed generative systems. The Syn., Sem., and HW columns encode evaluation coverage using compact labels rather than binary checkmarks, reflecting that semantic evaluation methods differ fundamentally across model families and are not directly interchangeable.

\begin{table}[!t]
\centering
\scriptsize

\begin{tblr}{
  width=\textwidth,
  colsep=1.5pt,
  rowsep=1pt,
  stretch=0.98,
  rows={valign=t},
  row{1}={font=\bfseries},
  cell{2-Z}{8}={font=\fontsize{6.0pt}{6.6pt}\selectfont},
  colspec={
    Q[l,wd=2.15cm]   
    Q[l,wd=1.40cm]   
    Q[l,wd=1.15cm]   
    Q[l,wd=1.10cm]   
    Q[c,wd=0.48cm]   
    Q[c,wd=0.60cm]   
    Q[c,wd=0.48cm]   
    X[l]             
  },
}
\toprule
System &
Family &
{Out-\\put} &
Size &
Syn. &
Sem. &
HW &
Evaluation Notes \\
\midrule

{QuantumGPT-\\124M\\\cite{hf_quantumgpt_124m,hf_quantum_circuits_8k}}
& {Spec.\\small LM}
& {QASM\\2.0}
& {124\,M}
& \checkmark
& {Lim}
& {---}
& Parser validation; manual inspection on $\leq$5 qubit circuits (no oracle) \\

{Granite3.2\-8b\\Qiskit\\\cite{hf_granite_qiskit,qiskit_humaneval}}
& {Qiskit\\LLM}
& {Qiskit\\(Py)}
& {8\,B}
& \checkmark
& {UT}
& {---}
& QiskitHumanEval unit tests (pass@$k$); coding benchmarks \\

{Qwen2.5\-14B\\Qiskit\\\cite{hf_qwen_qiskit,qiskit_humaneval}}
& {Qiskit\\LLM}
& {Qiskit\\(Py)}
& {14.7\,B}
& \checkmark
& {UT}
& {---}
& Similar unit-test-driven evaluation \\

{Agent-Q (SFT)\\\cite{agentq_paper,hf_agentq_sft,hf_graph_data_quantum_rl}}
& {Optim-\\LLM\\(SFT)}
& {QASM\\3.0}
& {Unspec.\\\textsuperscript{b}}
& \checkmark
& {DA}
& {---}
& Distribution and expectation-value alignment \\

{QUASAR\\(SFT+RL)\\\cite{quasar_paper,hf_quasar_rl,hf_graph_data_quantum_rl}}
& {Verifier\\RL}
& {QASM\\3.0}
& {4\,B}
& \checkmark
& {DA}
& {---}
& Hierarchical 4-level reward; pass@$k$ on syntax $+$ objective alignment \\

{genQC\\\cite{genqc_paper,gh_genqc,hf_genqc_unitary}}
& Diffusion
& {Circuit\\tok.}
& {Unspec.\\\textsuperscript{a}}
& \checkmark
& {PF}
& {---}
& Process fidelity; compilation metrics (3--5 qubits) \\

{KetGPT\\\cite{ketgpt_paper,gh_ketgpt}}
& {Trans.\\gen.}
& {QASM\\2.0}
& {Unspec.\\\textsuperscript{a}}
& \checkmark
& {RP}
& {---}
& Real-vs.-random classification + structural analysis (realism proxy, not task semantics) \\

{AltGraph\\\cite{altgraph_paper}}
& {Graph\\rewr.}
& {Circ.\\DAG}
& {Unspec.\\\textsuperscript{a}}
& \checkmark
& {PF}
& {3a}
& Density-matrix MSE; depth/gate reduction measured post-transpilation (L3a) \\

{Q-Fusion\\\cite{qfusion_paper}}
& {Graph\\diff.}
& {Circ.\\DAG}
& {Unspec.\\\textsuperscript{a}}
& \checkmark
& {Lim}
& {---}
& Validity rate in tested regimes; limited semantic eval \\

{UDiTQC\\\cite{uditqc_paper}}
& {Diff.\\transf.}
& {Circuit\\tok.}
& {Unspec.\\\textsuperscript{a}}
& \checkmark
& {PF}
& {---}
& Process fidelity on entanglement/compilation; outperforms genQC \\

{Barta et al.\\\cite{barta2025diffusion}}
& {Diff.\\(PQC)}
& {Param.\\circ.}
& {Unspec.\\\textsuperscript{a}}
& \checkmark
& {Lim}
& {---}
& Diffusion for parameterized circuits; QCE 2025 \\

{genQC v2\\\cite{genqc_v2}}
& {Diff.\\(multi)}
& {Param.\\circ.}
& {Unspec.\\\textsuperscript{a}}
& \checkmark
& {Lim}
& {---}
& Multimodal diffusion over discrete structure and continuous parameters; limited evaluation disclosure \\

{QAgent\\\cite{qagent_paper}}
& {Agentic\\LLM}
& {Open-\\QASM}
& {Base\\LLM}
& \checkmark
& {DA}
& {---}
& Multi-agent RAG+CoT; 71.6\% improvement over baselines \\

\bottomrule
\end{tblr}

{\fontsize{6.4pt}{7.0pt}\selectfont
\textbf{Evaluation layers}: Syn.\ = Syntactic validity (L1); Sem.\ = Semantic method (L2); HW = Hardware (L3).\\
\textbf{Semantic codes}: UT = unit tests; PF = process fidelity / density-matrix distance; DA = distributional $+$ expectation-value alignment; RP = realism proxy (structural similarity); Lim = limited or manual only.\\
\textbf{Hardware codes}: 3a = post-transpilation resource metrics reported; --- = no hardware-level evaluation.\\
\textsuperscript{a}Model size unspecified in source; architecture described qualitatively.\\
\textsuperscript{b}Paper text does not clearly specify the base-model size; the released Hugging Face implementation is 4\,B.
}
\caption{Reviewed generative systems with artifact types, training regimes, and evaluation coverage.}
\label{tab:models}
\end{table}

\FloatBarrier

\subsection{Supporting Datasets and Benchmarks}\label{sec:datasets}
Table~\ref{tab:datasets} summarizes key datasets and benchmarks that support the generative systems reviewed above. No single dataset currently addresses all three evaluation layers, and schema differences between OpenQASM 2.0 and 3.0 datasets remain a practical barrier to cross-system benchmarking. Benchmark suites such as QASMBench \cite{qasmbench_pnnl,qasmbench_repo} and QCircuitBench \cite{qcircuitbench_paper} are not generative models themselves but provide essential evaluation infrastructure: QASMBench is associated with execution fidelity measurements on real devices (IBM, IonQ, Rigetti), while QCircuitBench supplies 120,290 algorithm-design instances with automatic verification oracles spanning 25 algorithms in both OpenQASM 3.0 and Qiskit/Cirq formats.

\begin{table}[!htbp]
\centering
\scriptsize
\setlength{\tabcolsep}{2.5pt}
\renewcommand{\arraystretch}{1.10}
\begin{tabularx}{\textwidth}{%
  @{}%
  >{\RaggedRight\arraybackslash}p{3.0cm}%
  >{\RaggedRight\arraybackslash}p{2.3cm}%
  >{\RaggedRight\arraybackslash}p{1.2cm}%
  >{\RaggedRight\arraybackslash}X%
  @{}}
\toprule
\textbf{Dataset} &
\textbf{Primary use} &
\textbf{Scale} &
\textbf{Notes} \\
\midrule

quantum-circuits-8k \cite{hf_quantum_circuits_8k}
& Text$\rightarrow$OpenQASM 2.0 SFT
& $\sim$8\,k
& Synthetic with paraphrase augmentation; small-circuit emphasis \\[3pt]

graph-data-quantum-rl \cite{hf_graph_data_quantum_rl}
& Optimization-circuit generation and RL
& 14.5\,k rows
& Prompts, graphs, Hamiltonians, OpenQASM 3.0 circuits, solutions \\[3pt]

QASMBench \cite{qasmbench_repo,qasmbench_pnnl}
& OpenQASM-2 benchmark suite
& diverse
& Curated benchmark circuits and circuit-level metrics \\[3pt]

QCircuitBench \cite{qcircuitbench_paper}
& Algorithm design benchmarking
& 120,290
& QASM 3.0 + code (Qiskit/Cirq) + oracles / verification functions \\[3pt]

Quantum\-LLM\-Instruct \cite{quantumllminstruct_paper,hf_quantumllminstruct}
& Broad quantum instruction data
& 500k+ claimed
& Paper/model card claim 500k+ instruction-tuning pairs across 90+ quantum domains; current public HF viewer exposes 5.15k rows \\

\bottomrule
\end{tabularx}
\caption{Datasets and benchmarks supporting quantum circuit/code generation and evaluation.}
\label{tab:datasets}
\end{table}

\FloatBarrier

\section{Evaluation Framework}\label{sec:eval_stack}
Across model families, evaluation decomposes into three layers:
\begin{enumerate}[leftmargin=*]
  \item \textbf{Syntax}: parsing, compilation, or import success. For graph- and DAG-based generators, ``syntactic validity'' means structural well-formedness (valid DAG topology, legal gate placements) rather than parser-valid program text.
  \item \textbf{Semantics}: the method used to assess whether the generated artifact is \emph{correct}. This varies fundamentally across families: unit tests for code generation; process fidelity or density-matrix distances for compilation; expectation-value and distribution alignment for optimization tasks; and realism proxies (e.g., real-vs-random classification) for dataset augmentation. These methods are not interchangeable, and a system evaluated by one method cannot be directly ranked against a system evaluated by another.
  \item \textbf{Hardware/resources}, decomposed into two sublayers:
  \begin{enumerate}[label=3\alph*.,leftmargin=1.5em]
    \item \textbf{Compilability and resource realism}: transpilation to a target device's native gate set and connectivity succeeds; resulting circuit depth, SWAP count, and two-qubit gate overhead are acceptable.
    \item \textbf{Empirical execution}: the transpiled circuit is executed on a real QPU; measured output distributions are compared to ideal simulation using metrics such as Hellinger fidelity or total variation distance.
  \end{enumerate}
\end{enumerate}
This sublayer distinction is diagnostic: a system may address~3a (AltGraph measures post-transpilation depth and gate counts) without addressing~3b (no system in the reviewed corpus reports QPU execution results as part of model evaluation).

Benchmark suites such as QiskitHumanEval \cite{qiskit_humaneval} formalize unit-test evaluation for Qiskit code generation. Optimization-focused systems such as QUASAR emphasize simulator-driven objective metrics and pass@$k$ variants over multiple correctness criteria \cite{quasar_paper,hf_quasar_rl}.

\paragraph{Systematic application.}
Table~\ref{tab:models} applies this framework to all reviewed systems through the Sem. and HW columns. The pattern is clear: every reviewed system addresses Layer~1 (syntax), most address Layer~2 (semantics) to some degree, and \emph{none} addresses Layer~3b (empirical hardware execution). Only AltGraph partially addresses Layer~3a. This observation is elaborated in \S\ref{sec:hw_gap}.

\paragraph{Task-objective-to-evaluator mapping.}
A practitioner selecting a semantic evaluator must match the task objective to an appropriate metric. Table~\ref{tab:eval_map} provides concrete guidance.

\begin{table}[!htbp]
\centering
\scriptsize
\setlength{\tabcolsep}{3pt}
\renewcommand{\arraystretch}{1.15}
\begin{tabularx}{\textwidth}{%
  @{}%
  >{\RaggedRight\arraybackslash}p{2.8cm}%
  >{\RaggedRight\arraybackslash}p{3.5cm}%
  >{\RaggedRight\arraybackslash}X%
  @{}}
\toprule
\textbf{Task Objective} & \textbf{Recommended Evaluator} & \textbf{Known Failure Modes} \\
\midrule
Compilation to target unitary & Process fidelity, diamond norm proxy, or equivalence checking on a basis subset & Relative phase errors invisible to basis-restricted measurement; partial basis checking misses errors on untested inputs; ancilla garbage passes fidelity but fails full equivalence \\[3pt]
Optimization ansatz generation & Energy expectation value $+$ convergence speed $+$ robustness under re-optimization & Low-energy ansatz may be a local minimum; convergence speed conflated with initial parameter sensitivity \\[3pt]
Algorithm design tasks & Oracle-based functional checks (as in QCircuitBench \cite{qcircuitbench_paper}) & Oracle leakage if test structure correlates with training data; hard to verify beyond provided oracles \\[3pt]
Code assistants (Qiskit) & Unit tests $+$ execution traces (QiskitHumanEval \cite{qiskit_humaneval}) & Tests check observable behaviour, not internal correctness; ancilla state and relative phase may be ignored \\[3pt]
Dataset augmentation & Real-vs-random classification $+$ structural analysis (KetGPT \cite{ketgpt_paper}) & Distribution matching without semantic grounding; generated circuits may be syntactically realistic but computationally trivial \\
\bottomrule
\end{tabularx}
\caption{Mapping task objectives to recommended semantic evaluators and known failure modes.}
\label{tab:eval_map}
\end{table}

\paragraph{Metric gaming and composite evaluation.}
Any fixed evaluation metric is susceptible to gaming. Distribution-matching metrics can be satisfied by circuits that reproduce correct measurement statistics while implementing an incorrect unitary. Unit-test evaluation can be gamed by overfitting to test-case structure. Fidelity metrics are robust against such shortcuts but exponentially expensive at scale. These failure modes motivate \emph{composite} evaluation protocols combining metrics from different paradigms, as well as adversarial test suites targeting common evaluator blind spots.

\section{Hardware Gap and Transpilation}\label{sec:hw_gap}

\paragraph{Hardware evaluation as a field-wide gap.}
A substantive finding of this review is that \textbf{none of the thirteen reviewed generative systems reports end-to-end hardware execution results} (generation $\rightarrow$ transpile $\rightarrow$ execute $\rightarrow$ compare) as part of model evaluation. Layer~3b is absent from the generative model corpus. Benchmark suites such as QASMBench \cite{qasmbench_pnnl} are associated with hardware execution fidelity measurements, but QASMBench is evaluation infrastructure rather than a generative system. Among the generative systems, only AltGraph partially addresses Layer~3a by measuring post-transpilation depth and gate counts; no system closes the loop to Layer~3b.

\emph{The following protocol is proposed by this review as a direction for future evaluation practice; it is not established in the reviewed corpus.} A hardware evaluation protocol for generative quantum circuits might include: (a)~transpilation to a specific device's native gate set and connectivity (e.g., IBM Eagle 127-qubit heavy-hex), (b)~execution with multiple shot counts, (c)~comparison of measured distributions against ideal simulation using Hellinger fidelity or total variation distance, and (d)~resource accounting (SWAP insertions, final circuit depth, execution time relative to $T_1$/$T_2$ coherence~times).

\paragraph{Transpilation constraints.}
All generated quantum circuits must pass through transpilation before hardware execution. Transpilers such as the Qiskit transpiler \cite{qiskit_transpiler}, BQSKit \cite{bqskit}, and SABRE \cite{sabre_routing} perform gate decomposition into native gate sets, qubit routing, and optimization passes. In the public artifacts reviewed here, none of the generative systems explicitly accounts for hardware connectivity constraints or native gate sets during generation. Agent-Q, QUASAR, and QAgent generate circuits using abstract gate sets that assume all-to-all connectivity. genQC generates from a fixed gate pool that may not align with target hardware. AltGraph is closest to hardware awareness via post-transpilation metrics, but transpilation is applied \emph{after} generation rather than constrained \emph{during} generation.

This means generative models currently solve a subset of the full circuit design problem: they produce logically correct circuits that may require substantial transpilation overhead. \emph{We propose} that future systems incorporating transpilation constraints during generation---e.g., conditioning on device connectivity graphs or penalizing SWAP-heavy circuits during RL---would address a significant practical gap.

\paragraph{Minimal reporting baseline.} \emph{As a recommendation from this review}: even without transpilation-aware generation, a straightforward improvement would be to always report post-transpilation metrics under a standard reference backend, including at minimum: (i)~SWAP overhead ratio, (ii)~depth blow-up factor, and (iii)~the target backend topology. These require only a single transpiler call and would enable cross-system comparison on a hardware-realism dimension currently absent from all reviewed evaluations.

\section{Discussion}\label{sec:discussion}

\subsection{Evaluation Standardization}
The primary bottleneck remains comparability: unit-test pass rates, distributional alignment, and fidelity proxies are not directly interchangeable. Benchmarks may be gamed if they measure proxies rather than the task objective \cite{qiskit_humaneval,quasar_paper,genqc_paper}.

As a concrete illustration, consider a 5-qubit circuit evaluated by both pass@$k$ and process fidelity. A circuit implementing the correct computational-basis mapping (passing the unit test) may achieve $F < 0.8$ if it agrees on tested observable behaviour but differs in relative phases, ancilla state, or behaviour on untested inputs. Conversely, a circuit with $F = 0.99$ may fail a unit test that checks a side-effect (e.g., qubit ordering convention) the fidelity metric ignores. These divergences reflect structural differences between evaluation paradigms that prevent cross-system ranking.

\subsection{Data Provenance and Reproducibility}
Schema mismatches impede dataset reuse. The quantum-circuits-8k dataset uses OpenQASM 2.0 syntax, while graph-data-quantum-rl uses OpenQASM 3.0 with typed variables and parameterized gates. A model trained on one format cannot be directly evaluated on the other without a translation layer, and automated QASM~2.0$\rightarrow$3.0 conversion is not lossless \cite{hf_quantum_circuits_8k}.

\subsection{Scaling and Verification Cost}
Scaling beyond small-qubit compilation is constrained by classical verification cost. Statevector simulation requires $O(2^n)$ memory; full unitary reconstruction costs $O(4^n)$. For $n = 50$, statevector storage alone demands approximately 18~petabytes, and full unitary equivalence checking is doubly intractable. Tensor-network and stabilizer-rank methods offer partial relief for structured circuits but do not generalize to arbitrary unitaries. This simulation wall is a fundamental barrier to scaling verifier-in-the-loop training beyond the 30--50 qubit regime \cite{genqc_paper,quasar_paper,agentq_paper}.

\subsection{Future Evaluation Directions}
\emph{The following strategies are proposed by this review as future evaluation directions; they are not established practice in the reviewed corpus.}

The path-dependent semantics of OpenQASM 3.0 create evaluation challenges beyond those of 2.0's straight-line circuits. Three strategies merit consideration: (i)~\emph{bounded-path execution}---enumerate all classical branch paths up to a coverage bound and verify each path's unitary independently; (ii)~\emph{trace-based unit testing}---specify expected measurement and classical-variable traces for representative inputs; and (iii)~\emph{symbolic execution}---propagate symbolic states through classical branches to derive path conditions and verify equivalence on each feasible path. None of the reviewed systems currently employs these strategies.

\subsection{Threats to Validity}\label{sec:threats}
Several limitations should be considered when interpreting this review:
\begin{itemize}[leftmargin=*,nosep]
  \item \textbf{Single-reviewer process.} All screening, inclusion, and coding decisions were made by one author. No inter-rater reliability measure was computed. While appropriate for a scoping review of a small, emerging corpus, this introduces the possibility of systematic screening bias.
  \item \textbf{Corpus dependence on public disclosure.} The review is limited to systems with publicly available papers, model cards, or repositories. Closed-source industrial systems are excluded by design, which may omit significant work.
  \item \textbf{Provenance-based inclusion.} Three systems were identified through organization pages and citation tracing rather than keyword search. This reflects the limitations of keyword discovery in a fast-moving field but introduces discretionary inclusion that is not fully reproducible from the keyword protocol alone.
  \item \textbf{Mixed evidence quality.} Reviewed systems range from peer-reviewed publications to model cards with minimal documentation. Evaluation claims are taken at face value where replication was not feasible.
  \item \textbf{Non-comparable metrics.} The heterogeneity of evaluation methods across families means that cross-system ranking is not possible from the evidence base alone, despite the tabular presentation.
\end{itemize}

\section{Conclusion}
Generative AI for quantum circuits and code spans multiple model families unified by one central problem: enforcing semantic correctness under expensive verification. This review contributes a taxonomy grounded in artifact type $\times$ training regime, a three-layer evaluation framework (with Layer~3 decomposed into compilability and empirical execution sublayers) revealing that no reviewed generative model closes the loop to hardware execution, and a positioning against classical code generation that clarifies the unique challenges of the quantum setting. Future progress likely hinges on standardized evaluation protocols that separate syntax, semantics, and hardware realism; improved dataset provenance with attention to QASM version interoperability; transpilation-aware generation; and scalable verifier-in-the-loop methods that generalize beyond narrow problem families and small qubit counts.

\clearpage
\bibliographystyle{unsrt}
\bibliography{refs}

\end{document}